\documentclass[onecolumn,epjc3]{svjour3}

\RequirePackage{fix-cm}
\smartqed  
\RequirePackage{graphicx}
\RequirePackage{float}
\RequirePackage{hyperref}
\RequirePackage{amsmath}
\RequirePackage{amssymb}
\RequirePackage{mathtools}      

\journalname{Ind. J. Phys.}
\begin{document}

\title{Energy-Momentum Distribution in General Relativity for a Phantom
Black Hole Metric}


\author{P.K. Sahoo\thanksref{e1,addr1}
        \and
        I. Radinschi \thanksref{e2,addr2}
        \and
         K. L. Mahanta \thanksref{e3,addr3}    }


\thankstext{e1}{e-mail: pksahoo@hyderabad.bits-pilani.ac.in}
\thankstext{e2}{e-mail: radinschi@yahoo.com}
\thankstext{e3}{e-mail: kamal2\_m@yahoo.com}

\institute{Department of Mathematics, Birla Institute of
Technology and Science-Pilani, Hyderabad Campus, Hyderabad-500078,
India \label{addr1}
           \and
           Department of Physics, ``Gheorghe Asachi" Technical University, Iasi,
700050, Romania \label{addr2}
          \and 
Department of Mathematics, C.V. Raman College of Engineering, Bhubaneswar
752054,India \label{addr3}
}

\date{Received: date / Accepted: date}

\maketitle

\begin{abstract}
We use the M\o ller and Landau-Lifshitz energy-momentum definitions
in General Relativity (GR) to evaluate the energy-momentum
distribution of the phantom black hole space-time. The phantom black hole
model was applied to the supermassive black hole at the Galactic Centre. 
We obtain that in both pseudotensorial prescriptions the energy
distribution depends on the mass $M$ of the black hole, the phantom constant 
$p$ and the radial coordinate $r$. Further, all the calculated momenta are
found to be zero. The limiting cases $r\rightarrow 0$, $r\rightarrow \infty $
and $r\rightarrow -\infty $ have also been the subject of the study.


\end{abstract}


\section{Introduction}

\label{I}

The energy-momentum localization is one of the most important subjects which
remained unsolved in General Relativity (GR). Einstein was the first who
calculated the energy-momentum complex in a general relativistic system \cite%
{Ein1915}. The conservation law for the energy-momentum tensor i.e. matter
with non-gravitational fields for a physical system in (GR) is given by

\begin{equation}
\nabla _{\nu }T_{\mu }^{\nu }=0,  \label{eq1}
\end{equation}%
where $T_{\mu }^{\nu }$ is the symmetric energy-momentum tensor including
the matter and non-gravitational fields. The energy-momentum complex $\tau
_{\mu }^{\nu }$ satisfies a conservation law in the form of a divergence
given by 
\begin{equation}
\tau _{\mu ,\nu }^{\nu }=0,  \label{eq2}
\end{equation}%
with 
\begin{equation}
\tau _{\mu }^{\nu }=\sqrt{-g}(T_{\mu }^{\nu }+t_{\mu }^{\nu }),  \label{eq3}
\end{equation}%
where $g=\det g_{\nu \mu }$ and $t_{\mu }^{\nu }$ is the energy-momentum
pseudotensor of the gravitational field. The energy-momentum complex can be
written as 
\begin{equation}
\tau _{\mu }^{\nu }=\theta _{\mu ,\lambda }^{\nu \lambda },  \label{eq4}
\end{equation}%
with $\theta _{\mu }^{\nu \lambda }$ the superpotentials which are functions
of the metric tensor and its first derivatives. From the above discussion
the conclusion is that the energy-momentum complex is not uniquely defined.
This is because one can add a quantity with an identically vanishing
divergence to the expression $\tau _{\mu }^{\nu }$. Many famous physicists
like Tolman \cite{Tolman}, Landau and Lifshitz \cite{LL}, Papapetrou \cite%
{Papapetrou}, Bergmann \cite{BT} and Weinberg \cite{Weinberg} have given
different definitions for the energy-momentum complexes. These expressions
restricted the calculations of the energy distribution to quasi-Cartesian
coordinates. M\o ller \cite{Moller} introduced a new expression for the
energy-momentum complex which is consistent and enables one to perform the
calculations in any coordinate system. The M\o ller energy-momentum complex
is significant for describing the energy-momentum in (GR). In this regard,
we notice interesting results \cite{Many1,Many2} which recommend the M\o %
ller energy-momentum complex as an efficient tool for the energy-momentum
localization. Furthermore, the other energy-momentum complexes are also
important tools for the evaluation of the energy distribution and momentum
of a given space-time and yielded meaningful physical results. In the
context of the energy-momentum localization, it is very important to point
out the agreement between the Einstein, Landau-Lifshitz, Papapetrou,
Bergmann-Thomson, Weinberg and M\o ller definitions and the quasi-local mass
definition introduced by Penrose \cite{Penrose} and developed by Tod \cite%
{Tod} for some gravitating systems. The energy-momentum localization in a
Marder space-time has presented in \cite{Aygun/2007}. Further, the
energy-momentum distributions of texture and monopole topological defects
metrics in General Relativity are presented in \cite{Aygun/2010}.

In the early 90's, Virabhadra revived the issue of energy-momentum
localization by using different energy-momentum complexes in his pioneering
works \cite{Many3}. Rosen \cite{Rosen} employing the Einstein \ prescription
found that the total energy of the Friedman-Robertson-Walker (FRW)
space-time is zero. Johri et al. \cite{Johri} calculated the total energy of
the (FRW) universe in the Landau-Lifshitz prescription and found that is
zero at all times. The Einstein energy density for the Bianchi type-I
space-time is also zero everywhere \cite{BS}. Cooperstock and Israelit \cite%
{CI} evaluated the energy distribution for a closed universe and found the
zero value for a closed homogeneous isotropic( FRW) universe in (GR).
Further, \ to find an answer to the energy-momentum localization problem
several scientists have used various energy-momentum complexes to evaluate
the energy distribution for different space-times.

Moreover, recently the calculations performed for the $(3+1)$, $(2+1)$ and $%
(1+1)$ dimensional space-times have yielded physically reasonable results 
\cite{Many4,Many5,Many6,Many7}. We notice that several pseudotensorial
prescriptions give the same results for any metric of the Kerr-Schild class 
\cite{JMA}. Further, there is a similarity of some results with those
obtained by using the teleparallel gravity \cite{Many8}. Working with the
tetrad implies to encounter the notion of torsion, which can be used to
describe (GR) entirely with respect to torsion instead of curvature derived
from the metric only. This is called the teleparallel gravity equivalent to
(GR). Energy-momentum complexes are quasi-local quantities associated with a
closed 2-surface. Since Penrose introduced the definition of quasi-local
mass \cite{Penrose}, all the energy-momentum complexes present this
property. The issue of energy localization is also correlated with the
quasi-local energy given by Wang and Yau \cite{Wang/2009}. Searching for a
common quasi-local energy value represents in fact the reabilitation of the
pseudotensors. In this context, an important step has been made in the
energy localization research by Chen et. al. \cite{Chen/2018} who discovered
that with a 4D isometric Minkowski reference all of the quasi-local
expressions in a large class give the same energy-momentum

In this paper we use the M\o ller and Landau-Lifshitz prescriptions to
calculate the energy distribution for a metric that describes a phantom
black hole. There are two basic reasons to apply the M\o ller energy
momentum complex, the fact that it provides a powerful concept of energy and
momentum in General Relativity (GR) and that the calculations are not
restricted to quasi-Cartesian coordinates. Concerning the Landau-Lifshitz
energy-momentum complex, this is also an useful tool to calculate the energy
and momentum for a gravitating system and its use requires calculations to
be made in quasi-Cartesian coordinates. In this study, for the
Landau-Lifshitz prescription we have used the Schwarzschild Cartesian
coordinates $\{t$, $x$, $y$, $z\}$ and in the case of the M\o ller
prescription the Schwarzschild coordinates $\{t,$ $r,$ $\theta ,$ $\phi \}$,
respectively. The structure of the present paper is: in Section 2, we
describe the phantom black hole \cite{Ding} which is under study. Section 3
is focused on the presentation of the M\o ller energy-momentum complex and
of the calculations of the energy distribution and momenta of the phantom
black hole. In Section 4, we briefly introduce the Landau-Lifshitz
energy-momentum complex and we evaluate the expressions for the energy and
momenta. In Results and Discussion, we make a brief description of our
results and present some limiting cases. The Conclusions section is devoted
to the main conclusions of our study. In the paper, Greek (Latin) indices
run from $0$ to $3$ ($1$ to $3$) and we use geometrized units, i.e. $c=G=1$.

\section{Phantom Black Hole Metric}

\label{II}

The observation of very distant supernovae made with the Hubble Space
Telescope (HST) in 1998 indicated that the Universe is in an accelerated
expansion. The Universe is made up of 68 \% dark energy and the remaining
about 30 \% consists of dark matter and baryonic and nonbaryonic visible.
Dark energy can be described with the aid of a phantom scalar field that
represents a scalar with the minus sign for the kinetic term in the
Lagrangian. Nowadays, cosmological models with phantom scalar fields have
been extensively studied \cite{Many11}.

Furthermore, the phantom scalar field is of great interest in the physics of
black holes. The Lagrangian is given by 
\begin{equation}
L=\sqrt{-g}\biggl[-\frac{R}{8\pi G}+\epsilon g^{\mu \nu }\phi _{;\mu }\phi
_{;\nu }-2V(\phi )\biggr]
\end{equation}%
The structure of the Lagrangian includes a scalar field, the potential $%
V(\phi )$\ and $\epsilon $ that for the phantom takes the value $\epsilon
=-1 $.

A phantom black hole represents an exact solution of black holes in a
phantom field. The Bronnikov-Fabris phantom black hole metric \cite{BF},
later expressed by Ding et al. \cite{Ding} is given by

\begin{equation}
ds^{2}=f(r)dt^{2}-\frac{dr^{2}}{f(r)}-(r^{2}+p^{2})(d\theta ^{2}+\sin
^{2}\theta d\phi ^{2}),  \label{eq5}
\end{equation}

with 
\begin{equation}
f(r)=1-\frac{3M}{p}\biggl[\biggl(\frac{\pi }{2}-\arctan \frac{r}{p}\biggr)%
\biggl(1+\frac{r^{2}}{p^{2}}\biggr)-\frac{r}{p}\biggr],  \label{eq6}
\end{equation}

where $M$ is the mass parameter and $p$ is a positive constant relative to
the charge of phantom scalar fields known as the phantom constant \cite{Ding}
(here, $p$ is the phantom.) For the value $M=0$, \cite{Ding} the metric
describes the Ellis wormhole. The case $M<0$ corresponds to a wormhole which
is asymptotically flat at $r\rightarrow \infty $ and has an anti-de Sitter
bevaviour at $r\rightarrow -\infty $. For $M>0$ is obtained a regular black
hole that presents a Schwarzschild-like causal structure at large distances $%
r$.

The potential is given by

\begin{equation}
\frac{\phi}{\sqrt{2}}=\psi=\arctan\frac{r}{p}, \ \ V=\frac{3M}{p^3}\biggl[%
\bigg(\frac{\pi}{2}-\psi\bigg)(3-2\cos^2\psi)-3\sin\psi \cos\psi\biggr].
\end{equation}

The geometry of the phantom black hole can be used to obtain interesting
information concerning dark energy effects on strong gravitational lensing,
because the dark energy is modelled by the phantom scalar fields.

\section{M\o ller Energy-Momentum Complex in GR and the Energy Distribution
of the Phantom Black Hole}

\label{III}

\bigskip The energy-momentum complex of M{\o }ller \cite{Moller} is given by 
\begin{equation}
\mathcal{J}_{\nu }^{\mu }=\frac{1}{8\pi }\chi _{\nu \,\,,\,\lambda }^{\mu
\lambda },
\end{equation}%
where the anti-symmetric superpotentials $\chi _{\nu }^{\mu \lambda }$ are 
\begin{equation}
\chi _{\nu }^{\mu \lambda }=\sqrt{-g}\left( \frac{\partial g_{\nu \sigma }}{%
\partial x^{\kappa }}-\frac{\partial g_{\nu \kappa }}{\partial x^{\sigma }}%
\right) g^{\mu \kappa }g^{\lambda \sigma }
\end{equation}%
and satisfy the antisymmetric property 
\begin{equation}
\chi _{\nu }^{\mu \lambda }=-\chi _{\nu }^{\lambda \mu }.
\end{equation}
M\o ller's energy-momentum complex, like other energy-momentum complexes
satisfies the local conservation law 
\begin{equation}
\frac{\partial \mathcal{J}_{\nu }^{\mu }}{\partial x^{\mu }}=0,
\end{equation}%
where $\mathcal{J}_{0}^{0}$ and $\mathcal{J}_{i}^{0}$ are the energy and the
momentum densities, respectively. In the M{\o }ller definition, the
energy-momentum is given by 
\begin{equation}
P_{\nu }=\iiint \mathcal{J}_{\nu }^{0}dx^{1}dx^{2}dx^{3}.  \label{eq13}
\end{equation}

\bigskip The energy of the physical system has the following expression 
\begin{equation}
E=\int \int \int \Im _{0}^{0}dx^{1}dx^{2}dx^{3}.
\end{equation}

\bigskip Further, using Gauss's theorem, the energy $E$ can be written as 
\begin{equation}
E=\frac{1}{8\pi }\int \int \chi _{0}^{0i}n_{i}dS,  \label{eq15}
\end{equation}%
where $n_{i}$ is the outward unit normal vector over an infinitesimal
surface $dS$.\newline
The expression of the determinant of the metric (\ref{eq5}) is $%
g=-(r^{2}+p^{2})^{2}\sin ^{2}\theta $. The non-vanishing covariant
components of the metric (\ref{eq5}) are 
\begin{eqnarray}
g_{00} &=&f(r),  \nonumber \\
g_{11} &=&-\frac{1}{f(r)},  \nonumber \\
g_{22} &=&-(r^{2}+p^{2}),  \nonumber \\
g_{33} &=&-(r^{2}+p^{2})\sin ^{2}\theta .  \nonumber \\
&&
\end{eqnarray}

The corresponding contravariant components of the metric tensor are given by 
%
\begin{eqnarray}
g^{00} &=&\frac{1}{f(r)},  \nonumber \\
g^{11} &=&-f(r),  \nonumber \\
g^{22} &=&\frac{-1}{(r^{2}+p^{2})},  \nonumber \\
g^{33} &=&\frac{-1}{(r^{2}+p^{2})\sin ^{2}\theta },  \nonumber \\
&&
\end{eqnarray}

For the line element (\ref{eq5}) under consideration the only non-zero
superpotential is given by 
\begin{equation}
\chi _{0}^{01}=\frac{3M}{p}(r^{2}+p^{2})\biggl[\biggl(\arctan \frac{r}{p}-%
\frac{\pi }{2}\biggr)\frac{2r}{p^{2}}+\frac{2}{p}\biggr]\sin \theta .
\label{eq16}
\end{equation}%
Using the above expression and (\ref{eq15}) we obtain the energy
distribution of the phantom black hole 
\begin{equation}
E_{M}(r)=\frac{3Mr^{2}}{2p}(r^{2}+p^{2})\biggl[\biggl(\arctan \frac{r}{p}-%
\frac{\pi }{2}\biggr)\frac{2r}{p^{2}}+\frac{2}{p}\biggr].  \label{eq17}
\end{equation}%
Further,\ with (\ref{eq5}) and (\ref{eq13}) we found that all the momenta
vanish 
\begin{equation}
P_{r}=P_{\theta }=P_{\phi }=0.  \label{eq18}
\end{equation}

\begin{figure}[tbp]
\centering
\includegraphics[scale=0.4]{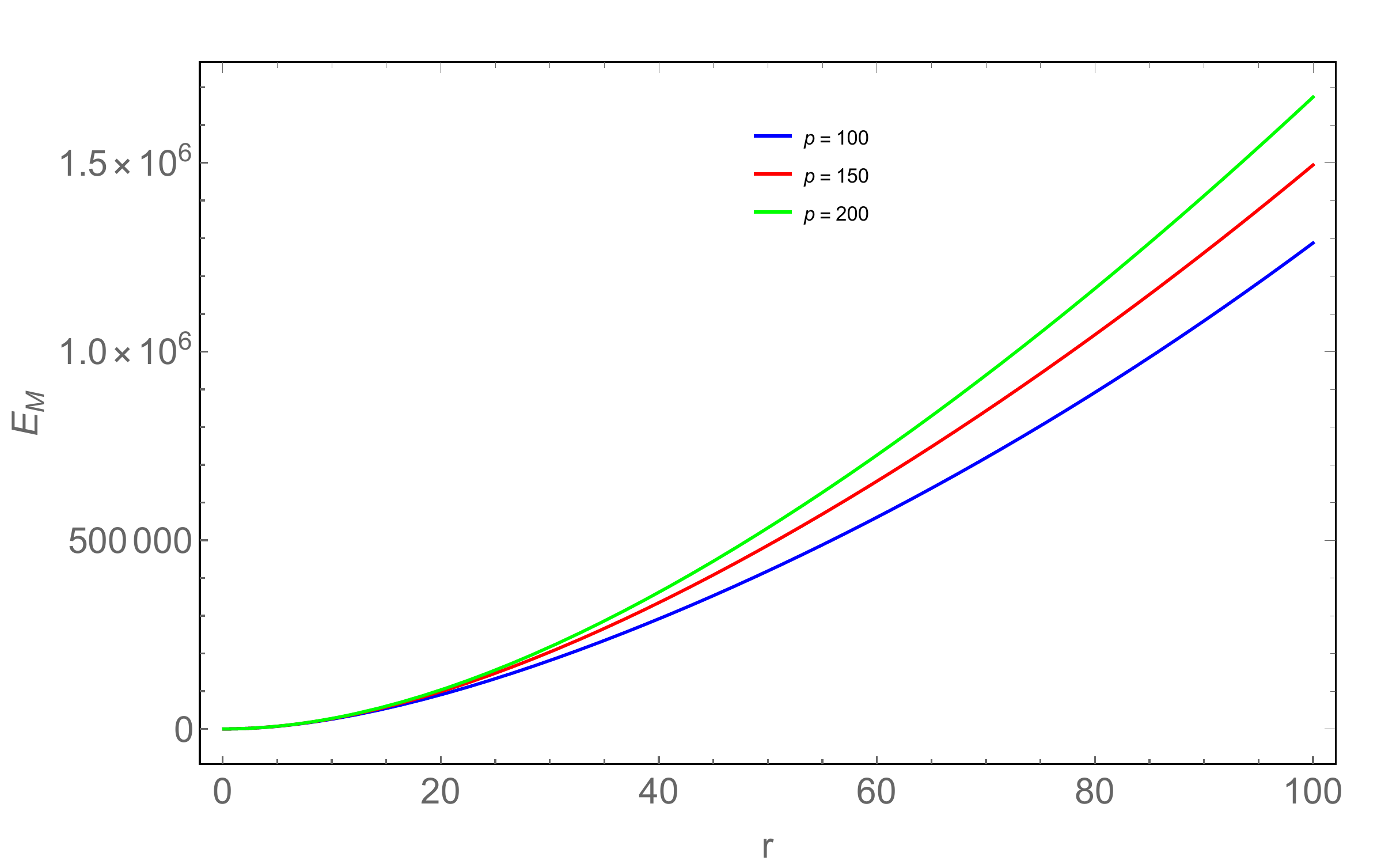}
\caption{The energy $E_{M}$ vs. the radial distance $r$ for several values
of phantom constant $p$ with $M=100$.}
\label{fig1}
\end{figure}

\begin{figure}[tbp]
\centering
\includegraphics[scale=0.4]{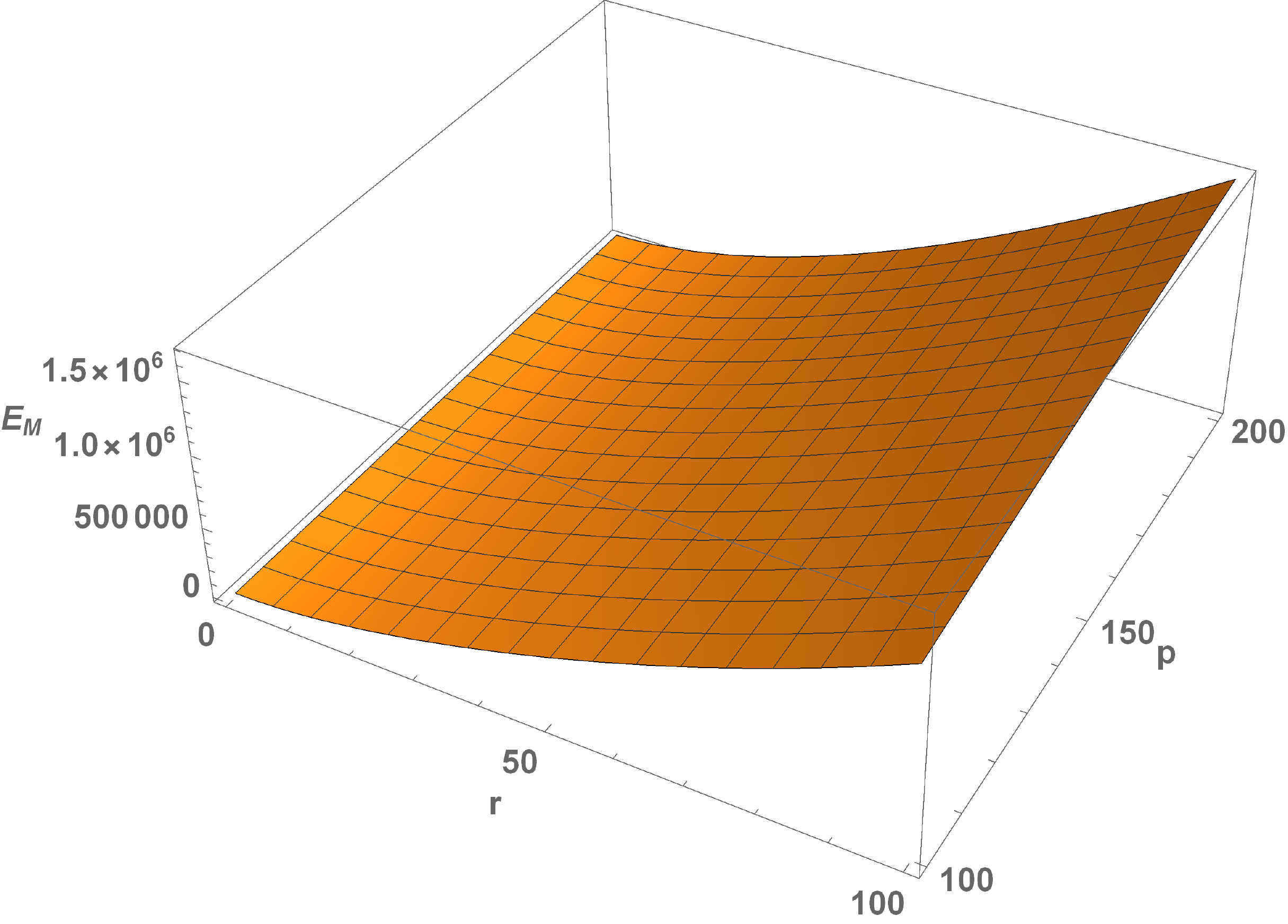}
\caption{The 2-dimensional surface plot of the energy $E_{M}$. $E_{M}$ is
plotted against the radial distance $r$ and the phantom constant $p$ with $%
M=100$.}
\label{fig2}
\end{figure}

We have plotted Fig. \ref{fig1} and Fig. \ref{fig2} to study the behaviour
of the energy distribution $E_{M}(r)$ \ when increasing the radial distance $%
r$ and the phantom constant $p$. In both figures we have fixed the mass
parameter $M$. From both graphical representations we notice that if $%
r\rightarrow 0$, $E_{M}(r)\rightarrow 0$ \ and when $r\rightarrow \infty $, $%
E_{M}(r)\rightarrow \infty $.

In Fig. 3 and Fig. 4 we plot the energy $E_{M}$ near zero. 
\begin{figure}[tbp]
\centering
\includegraphics[scale=0.4]{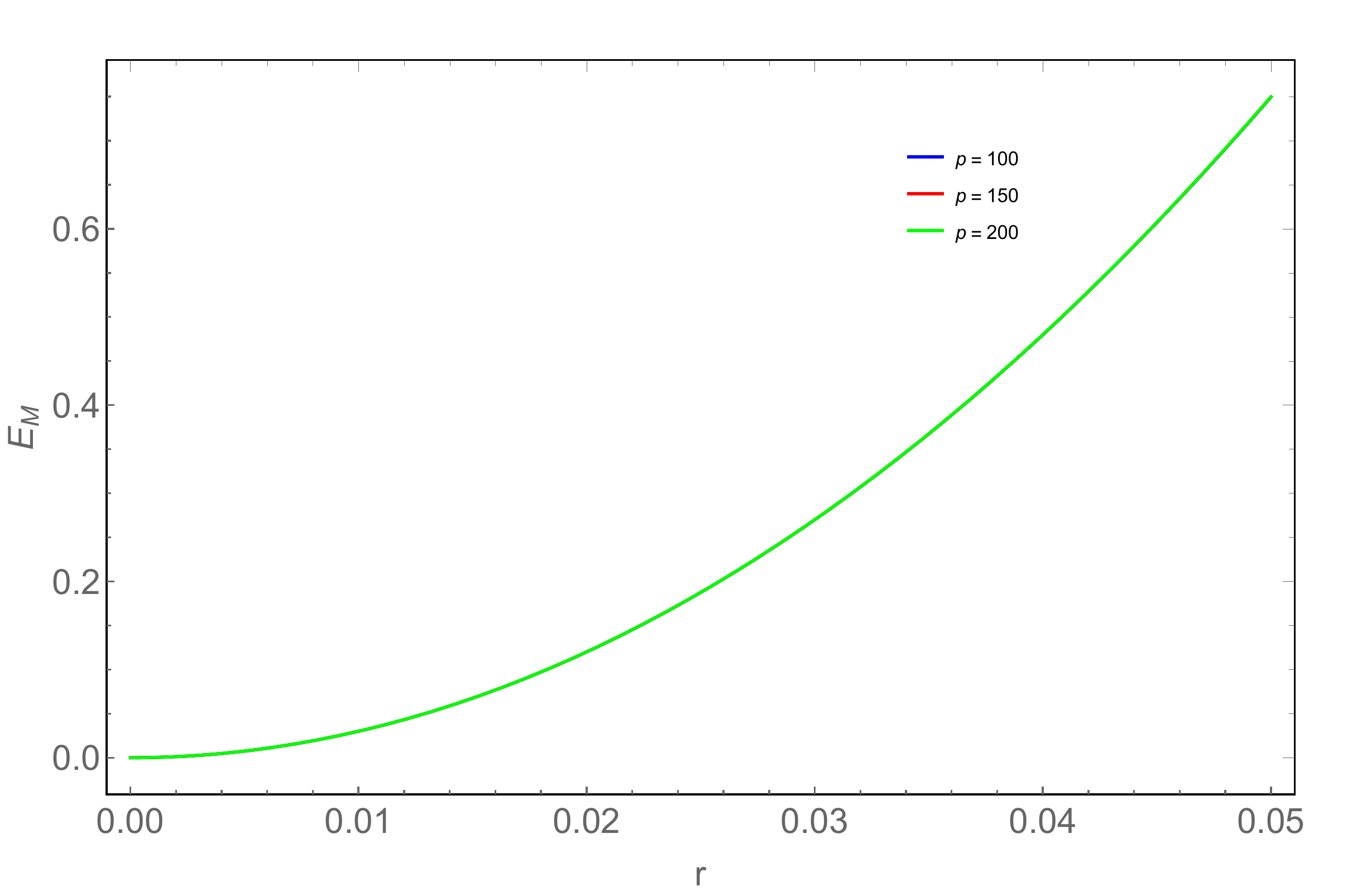}
\caption{The energy $E_{M}$ vs. the radial distance $r$ for several values
of phantom constant $p$ with $M=100$ near zero.}
\label{fig3}
\end{figure}
\begin{figure}[tbp]
\centering
\includegraphics[scale=0.4]{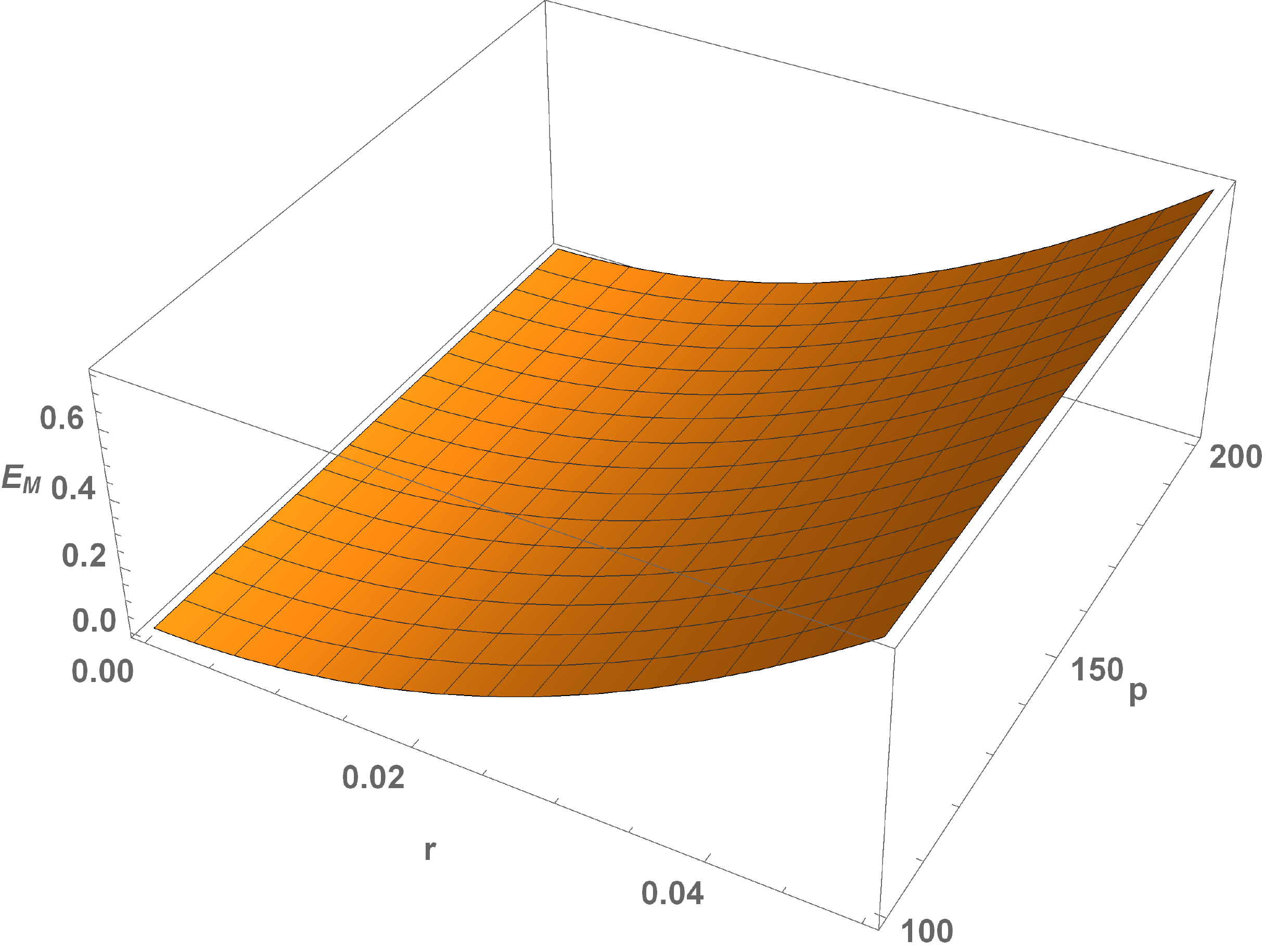}
\caption{The 2-dimensional surface plot of the energy $E_{M}$. $E_{M}$ is
plotted against the radial distance $r$ and the phantom constant $p$ with $%
M=100$ near zero.}
\label{fig4}
\end{figure}

\section{Landau-Lifshitz Energy-Momentum Complex in GR and the Energy
Distribution of the Phantom Black Hole}

\label{IV}

To perform the calculations of the energy distribution and momentum, the
line element (\ref{eq5}) is transformed to quasi-Cartesian coordinates $%
t,x,y,z$ using 
\begin{eqnarray}
x &=&r\sin \theta \cos \phi ,  \nonumber \\
y &=&r\sin \theta \sin \phi ,  \nonumber \\
z &=&r\cos \theta .
\end{eqnarray}%
For the line element (\ref{eq5}) we obtain the form 
\begin{eqnarray}
ds^{2} &=&f(r)dt^{2}-\frac{r^{2}+p^{2}}{r^{2}}(dx^{2}+dy^{2}+dz^{2}) 
\nonumber  \label{eq6a} \\
&&-\biggl(\frac{1}{f(r)}-\frac{r^{2}+p^{2}}{r^{2}}\biggr)\biggl(\frac{%
xdx+ydy+zdz}{r}\biggr)^{2},
\end{eqnarray}%
where 
\begin{equation}
r=\sqrt{x^{2}+y^{2}+z^{2}}.  \label{eq7}
\end{equation}%
The generalized Landau-Lifshitz energy-momentum complex for GR theory is
given by, \cite{LL} 
\begin{equation}
L^{\mu \nu }=\frac{1}{16\pi }S_{,\,\rho \sigma }^{\mu \nu \rho \sigma },
\label{eq19}
\end{equation}%
where the Landau-Lifshitz superpotentials are given by the expression 
\begin{equation}
S^{\mu \nu \rho \sigma }=-g(g^{\mu \nu }g^{\rho \sigma }-g^{\mu \rho }g^{\nu
\sigma }).  \label{eq20}
\end{equation}%
The $L^{00}$ and $L^{0i}$ components are the energy and the momentum
densities, respectively. In the Landau-Lifshitz prescription the local
conservation is respected 
\begin{equation}
L_{,\,\nu }^{\mu \nu }=0.  \label{eq21}
\end{equation}%
By integrating $L^{\mu \nu }$ over the 3-space one gets the following
expression for the energy and momentum 
\begin{equation}
P^{\mu }=\iiint L^{\mu 0}\,dx^{1}dx^{2}dx^{3}.  \label{eq22}
\end{equation}%
By using Gauss' theorem we obtain 
\begin{equation}
P^{\mu }=\frac{1}{16\pi }\iint S_{,\nu }^{\mu 0i\nu }n_{i}dS=\frac{1}{16\pi }%
\iint U^{\mu 0i}n_{i}dS.  \label{eq23}
\end{equation}

Using (\ref{eq19}), (\ref{eq23}) and the non-vanishining components of the
Landau-Lifshitz superpotentials, we obtain the energy distribution of the
phantom black hole 

\begin{equation}
E_{LL}(r)=-\frac{\left(p^2+r^2\right) \left(6 M \left(p^4-r^4\right) \tan
^{-1}\left(\frac{r}{p}\right)-3 \pi M p^4+6 M p^3 r-6 M p r^3+3 \pi M r^4+2
p^5\right)}{2 r^3 \left(-6 M \left(p^2+r^2\right) \tan ^{-1}\left(\frac{r}{p}%
\right)+3 \pi M p^2-6 M p r+3 \pi M r^2-2 p^3\right)}  \label{eq24}
\end{equation}

In this prescription also all the momenta vanish.

\[
P^{x}=P^{y}=P^{z}=0. 
\]

Fig. 5 and Fig. 6 show the dependence of the energy $E_{LL}$ on the radial
distance $r$ and phantom parameter $p$ for a constant value of the mass $M$
of the phantom black hole. 
\begin{figure}[]
\centering
\includegraphics[scale=0.4]{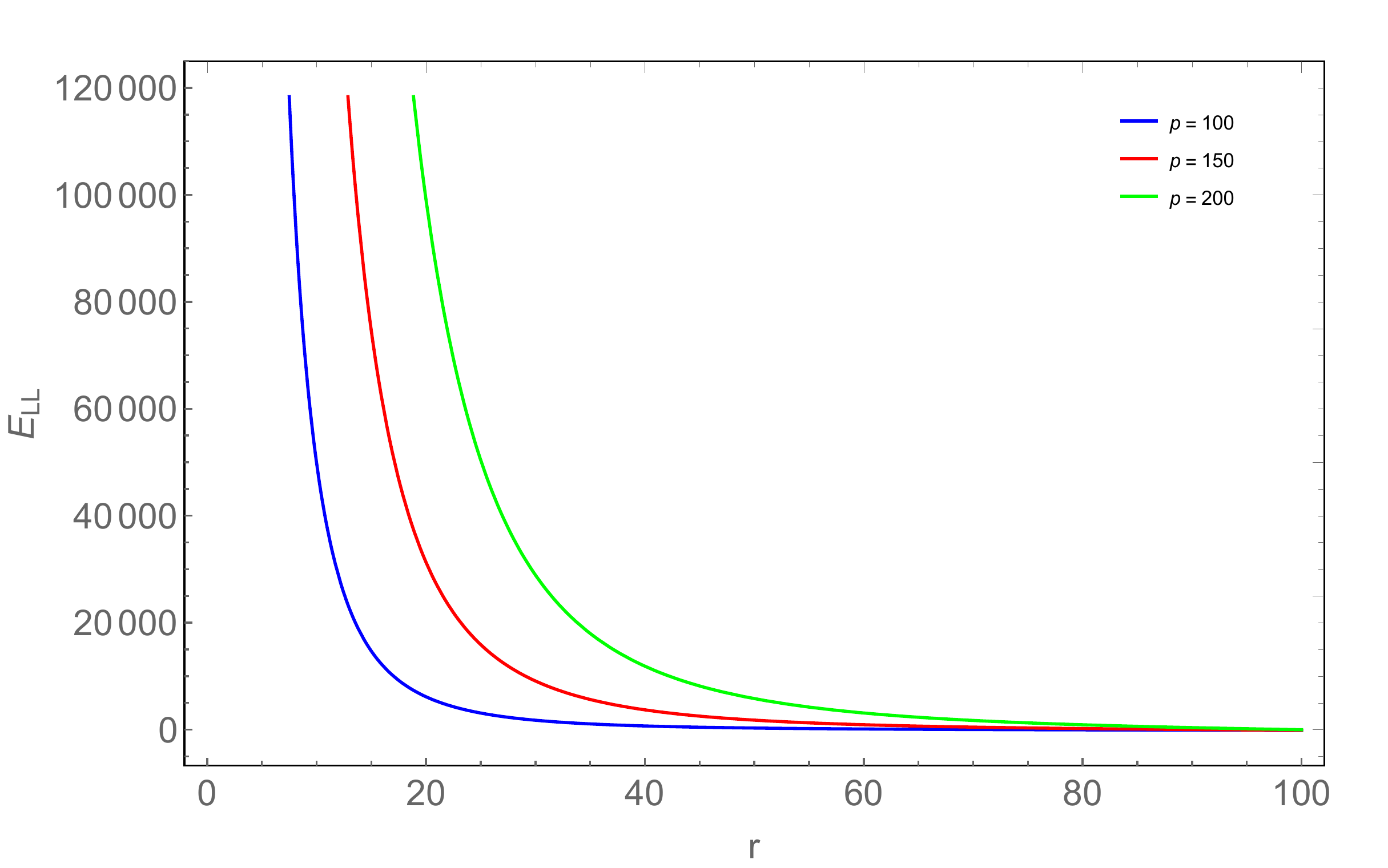}
\caption{The energy $E_{LL}$ vs. the radial distance $r$ for several values
of phantom constant $p$ with $M=100$.}
\label{fig5}
\end{figure}
\begin{figure}[]
\centering
\includegraphics[scale=0.4]{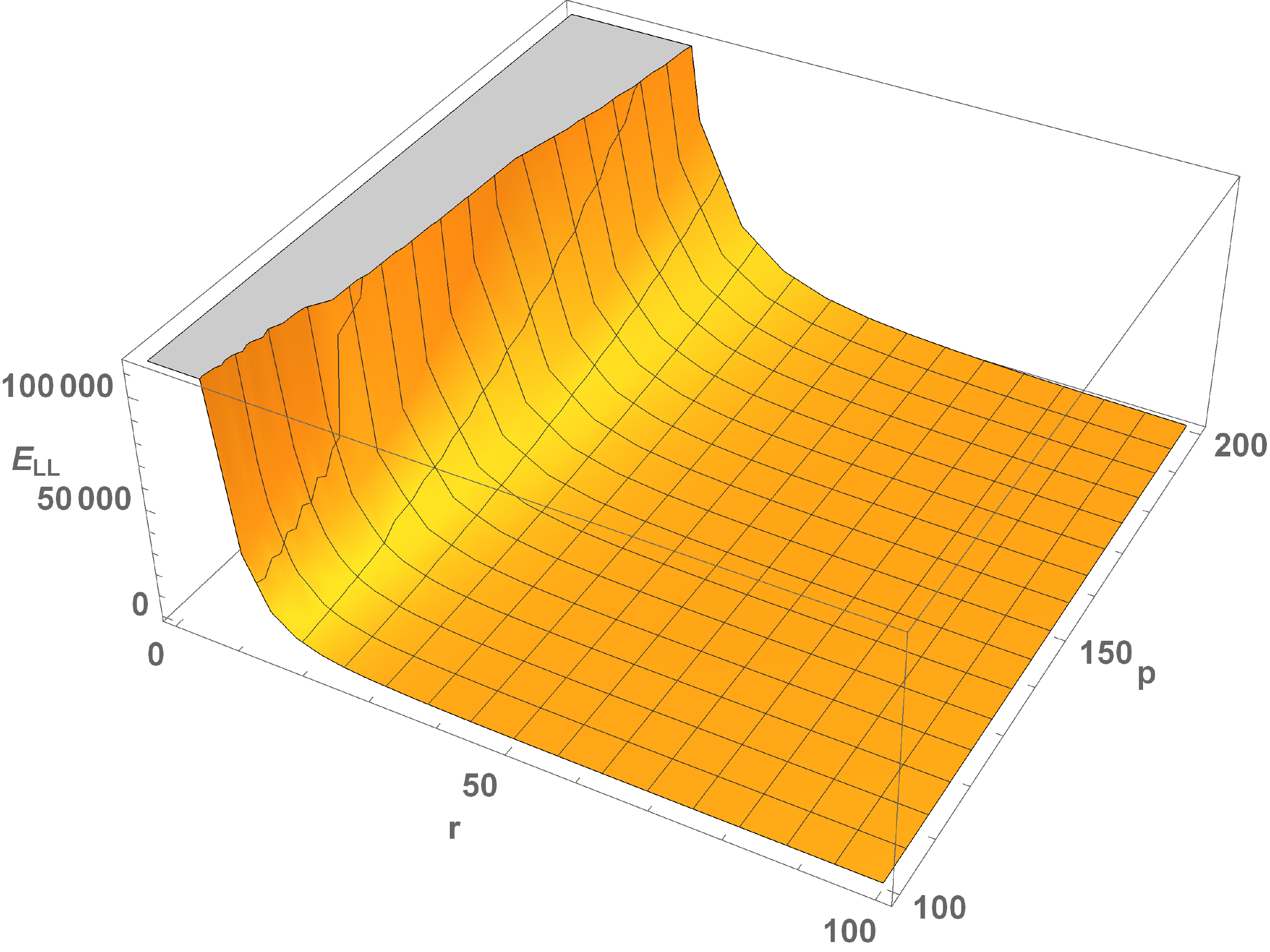}
\caption{The 2-dimensional surface plot of the energy $E_{LL}$. $E_{LL}$ is
plotted against the radial distance $r$ and the phantom constant $p$ with $%
M=100$.}
\label{fig6}
\end{figure}

The behaviour of the energy near zero is presented in Fig. 7 and Fig. 8.

\begin{figure}[]
\centering
\includegraphics[scale=0.4]{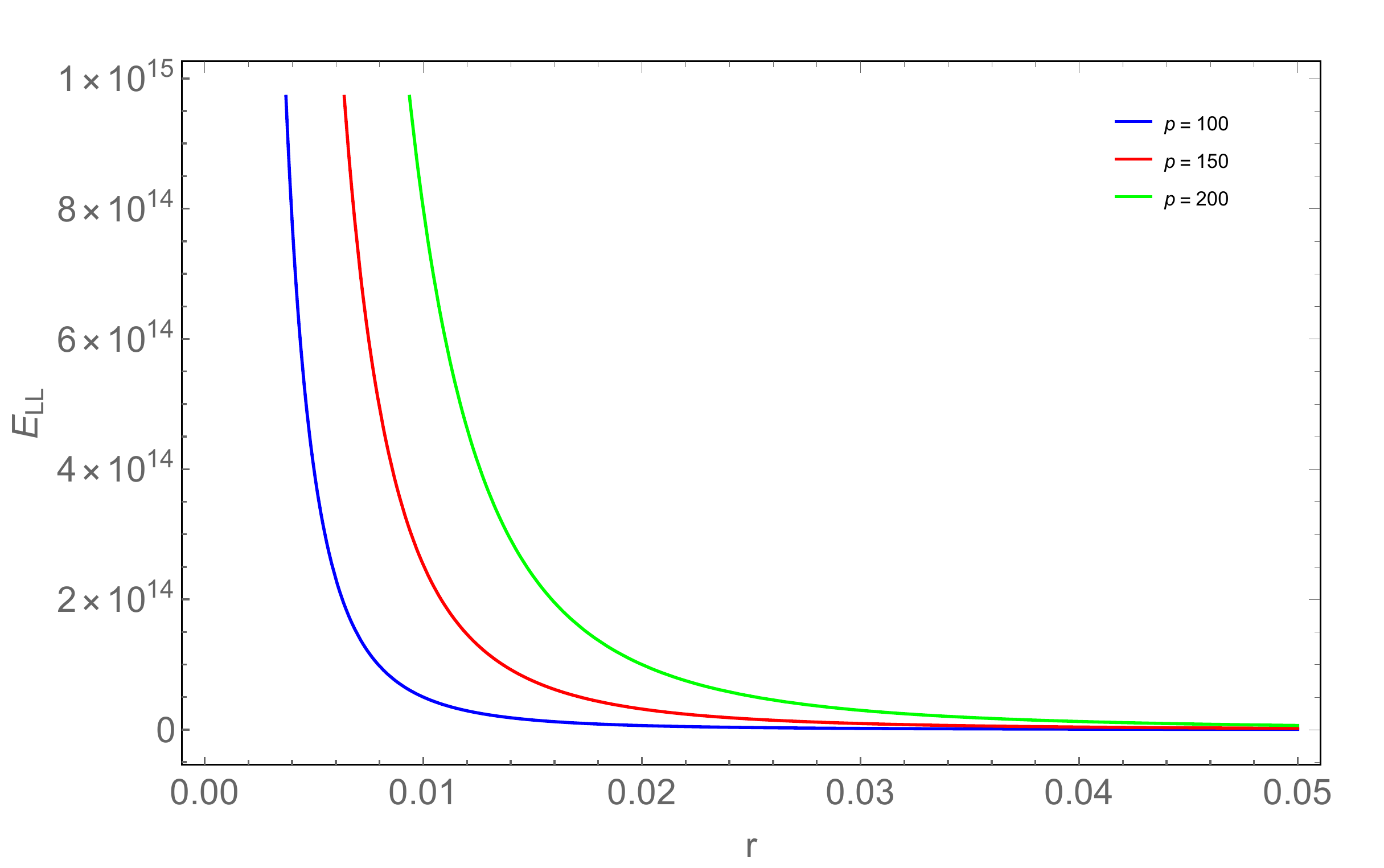}
\caption{The energy $E_{LL}$ vs. the radial distance $r$ for several values
of phantom constant $p$ with $M=100$ near zero.}
\label{fig7}
\end{figure}
\begin{figure}[]
\centering
\includegraphics[scale=0.4]{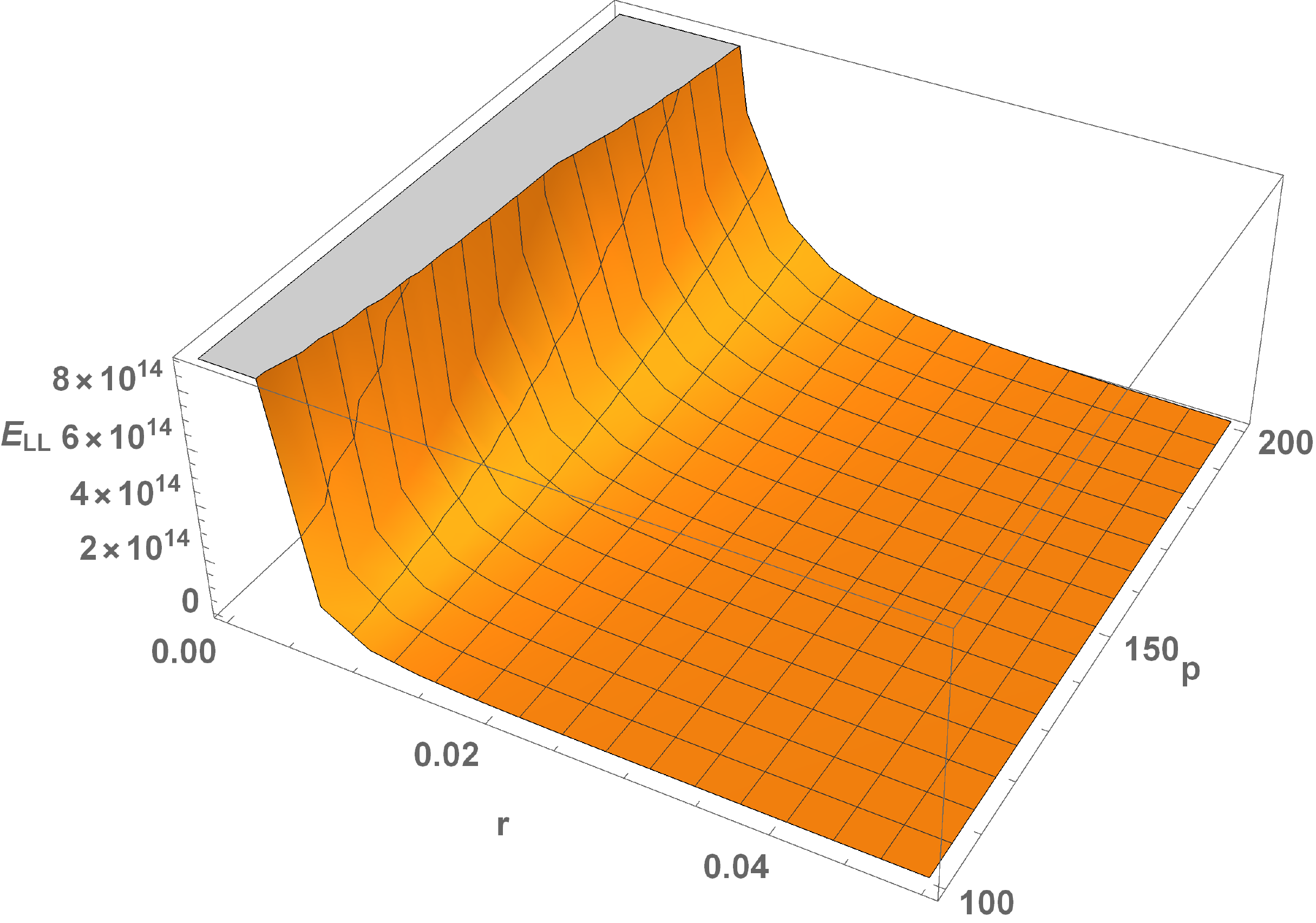}
\caption{The 2-dimensional surface plot of the energy $E_{LL}$. $E_{LL}$ is
plotted against the radial distance $r$ and the phantom constant $p$ with $%
M=100$ near zero.}
\label{fig8}
\end{figure}

\section{Results and Discussion}

\label{VI}

The energy-momentum complexes provide the same energy-momentum distribution
for many gravitating systems. However, for some space-times the results
obtained with various prescriptions differ each other. Hence, the debate on
the localization of energy \ is one of the most actual and interesting
problem in (GR). The study of the energy-momentum distribution can also give
a clear idea about the space-time. One can study the gravitational lensing
of the spacetime analyzing the energy. Virbhadra \cite{VirEllisLens} derived
interesting lensing phenomena using the analysis of the energy distribution
in curved space-time. We also point out some meaningful results obtained by
the authors with the pseudotensorial prescriptions, and in this view we draw
attention to some research papers elaborated in the last two decades on this
issue \cite%
{Radinschi/2001,Radinschi/2007,Radinschi/2010,Radinschi/2011,sahoo/2015,Radinschi/2016}%
.

In this paper we calculated the energy distribution of the phantom black
hole using the M\o ller and Landau-Lifshitz energy-momentum complexes. In
both prescriptions the energy distribution depends on the mass $M$ of the
black hole, the phantom constant $p$ and the radial coordinate $r$. $%
E_{M}(r) $ and $E_{LL}(r)$ represent the total (matter plus gravitational
field) energy within radius $r$ in the M\o ller and Landau-Lifshitz
prescriptions, respectively.\ From the calculations, it results that in
these prescriptions all the momenta are zero. The expressions of the energy
distribution obtained in these two prescriptions are different because of
the differences between the superpotentials which constitute the M\o ller
and Landau-Lifshitz energy-momentum complexes, as well as by the structure
of the studied metric. Furthermore, the difference in the order of magnitude
for the energy calculated in both prescriptions is due to different
expressions of the M\o ller and Landau-Lifshitz energy-momentum complexes.

A clarification and a physical interpretation of the results obtained with
the M\o ller and Landau-Lifshitz energy-momentum complexes is now needed. To
calculate the energy-momentum distributions it is important, because we can
obtain useful information about the gravitational background like the value
of the effective gravitational mass of the source of spacetime curvature,
and also a prediction about the gravitational lensing. The positive energy
distribution region plays the role of a convergent lens and the negative one
serves as a divergent lens.

A limiting case that is of special interest is the behaviour of the energy
near the origin, that is for $r\rightarrow 0$. For some spacetimes here the
metric goes infinite, a singularity arises and the energy distribution and
momentum deal with extreme values. This behavior is in connection with the
spacetime geometry. We expect that the expression of the energy distribution
will give us some details about the utility of the applied energy-momentum
complex. In the case of the M\o ller energy-momentum complex we found that
for $r\rightarrow 0$ the energy tends to zero. For the Landau-Lifshitz
energy momentum-complex for $r\rightarrow 0$ the energy tends to plus
infinity.

Also, the results obtained in this work exhibit that there is no finite
value for the energy when $r\rightarrow \infty $. One can compare these
results with our previous result \cite{sahoo/2015} obtained with the
Einstein energy momentum complex for the same metric. In our previous paper
the energy distribution is positive and becomes constant for increasing the
radial distance $r$. In the present work, for increasing the radial distance
the energy distribution becomes infinite for larger values of $r$ in the
case of the M\o ller prescription and tends to minus infinity in the case of
the Landau-Lifshitz prescription. The energy distribution calculated in the M%
\o ller prescription takes only positive values for any values of the
parameters $p$ and $M$. Furthermore, from our study we have detected that
the energy distribution in the Landau-Lifshitz prescription has both
positive and negative values for some preferred values of the parameters $p$
and $M$. These results come to support the use of the M\o ller and
Landau-Lifshitz energy-momentum complexes for the evaluation of the
energy-distribution of a given space-time, because the positive energy
region serves as a convergent lens and the negative one as a divergent lens 
\cite{Virbhadra/2000}.

The limiting cases $r\rightarrow 0$, $r\rightarrow \infty $ and $%
r\rightarrow -\infty $ are presented in Table 1.

\begin{table}[]
\caption{ }
\begin{center}
\begin{tabular}{|c|c|c|c|}
\hline
Energy & $r\rightarrow 0$ & $r\rightarrow \infty $ & $r\rightarrow -\infty $
\\ 
$E_{M}$ & $0$ & $\infty $ & $\infty $ \\ 
$E_{LL}$ & $\infty $ & $-\infty $ & $\infty $ \\ \hline
\end{tabular}%
\end{center}
\end{table}

The phantom scalar could play an important role in black hole physics and it
would be of interest to test this phantom field, the best approach being
gravitational lensing. Microlensing is useful to probe dark matter and dark
energy in the Galatic halo \cite{Chang/1979}. For the metric given by (\ref%
{eq5}) for a small value of the phantom $p$ or in the case $p\rightarrow 0$
the phantom black hole behaves as a so called Schwarzschild phantom black
hole, and the single event horizon is $r_{+}=2\,M$. If $p$ increases, the
radius of the event horizon decreases and a stronger effect from dark energy
is noticeable$.$The phantom $p$ has a behaviour similar to the electric
charge in the Reissner-Nordstr\"{o}m black hole allowing the comparison
between the phantom black hole lensing and the Reissner-Nordstr\"{o}m
lensing. So, the metric described by (\ref{eq5}) can yield useful
information about dark energy effects on strong gravitational lensing. Very
interesting is the behaviour of the energy distribution in the M\o ller and
Landau-Lifshitz prescriptions near the event horizon. As we pointed out, the
positive and negative regions of the energy serve as convergent and
divergent lenses, respectively. Both expressions of the energy distribution $%
E_{M}(r)$ and $E_{LL}(r)$ contain the phantom $p$ and this could have
effects on strong gravitational lensing. To study the behaviour of the
energy distribution near the event horizon we performed a Taylor expansion
of $E_{M}(r)$ and $E_{LL}(r)$ in function of $r=2\,M$ in the particular case 
$p=150$ and $M=100$, and we plot these expressions for $r=2\,M$ in Fig. 9
and Fig. 10, respectively. As we expected, the energy in the M\o ller
prescription near the event horizon $E_{HM}$ takes only positive values and
plays the role of a convergent lens. In the case of the Landau-Lifshitz
prescription, the energy near the event horizon $E_{HLL\text{ }}$takes both
positive and negative values and serves as a convergent and divergent lens,
respectively. Obviously, a deeper study of the effects of dark energy on the
strong gravitational lensing in the case of the phantom black hole is
required.

\begin{figure}[]
\centering
\includegraphics[scale=0.4]{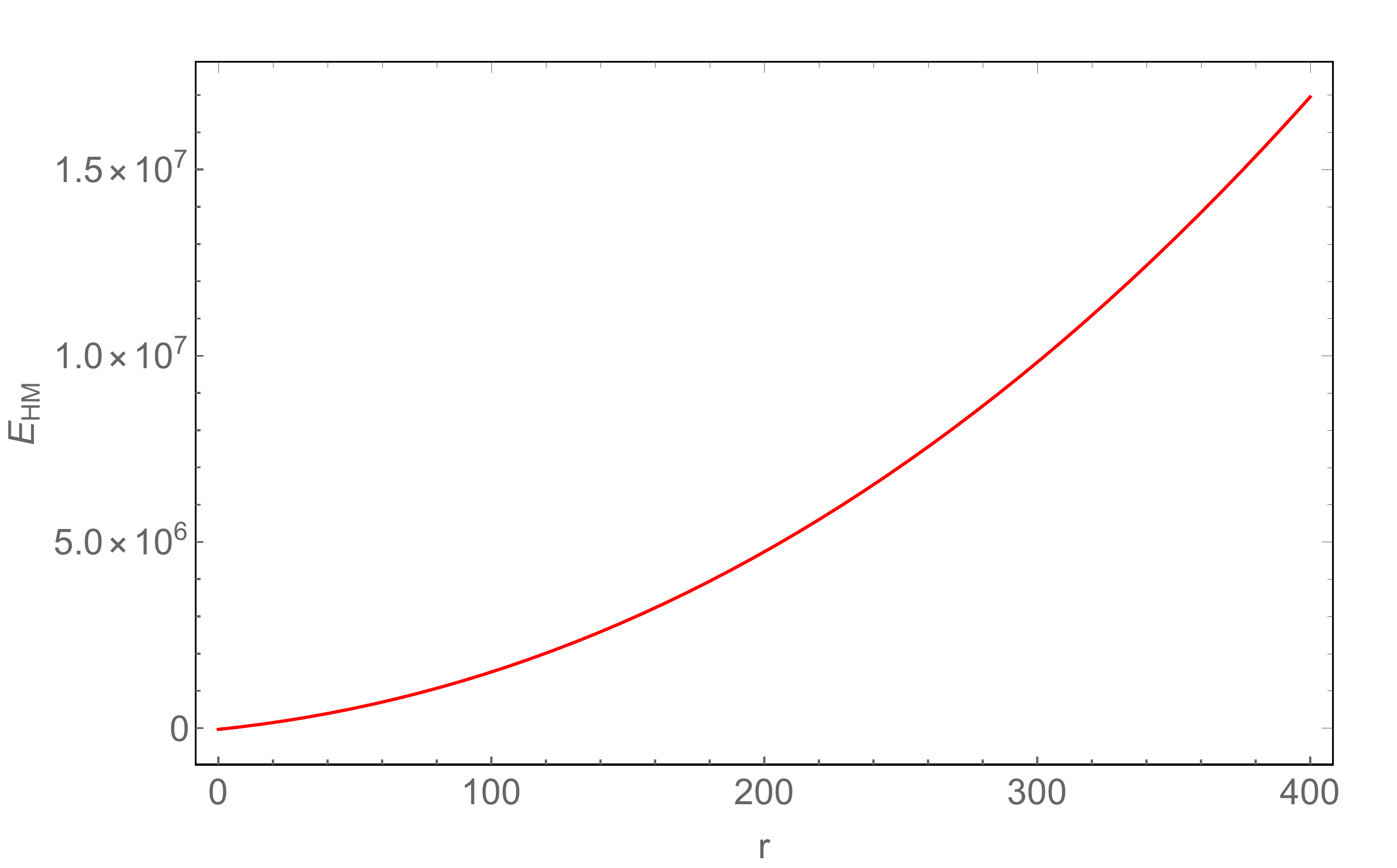}
\caption{The energy $E_{HM}$ near event horizon vs. the radial distance $r$
with $p=150$ and $M=100$.}
\label{fig9}
\end{figure}

\begin{figure}[]
\centering
\includegraphics[scale=0.4]{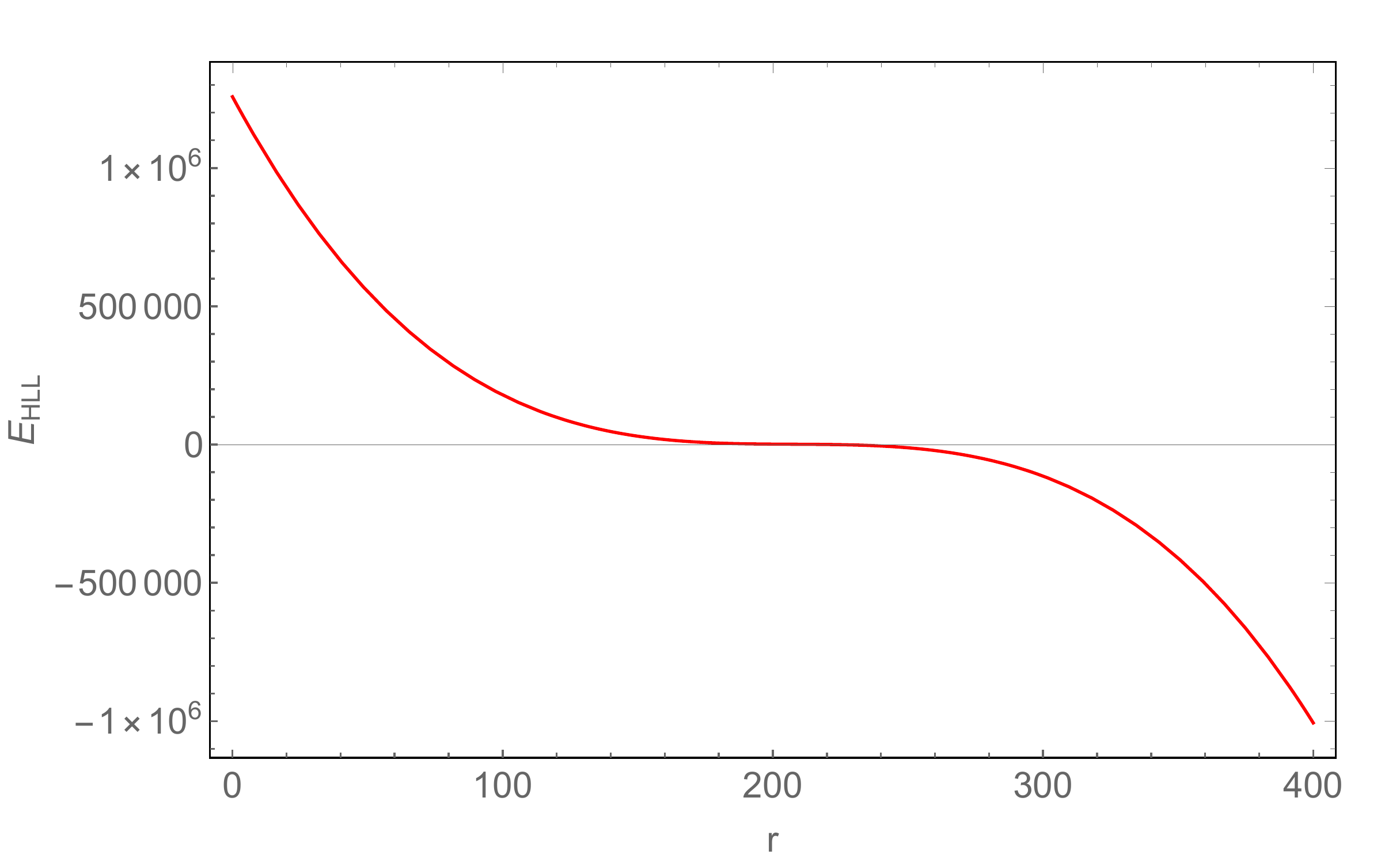}
\caption{The energy $E_{HLL}$ near event horizon vs. the radial distance $r$
with $p=150$ and $M=100$.}
\label{fig10}
\end{figure}

As stated by the results obtained in this work and in our previous work \cite%
{sahoo/2015} we can conclude that the Einstein and M\o ller prescriptions
are useful tools for the localization of energy. 


\section{Conclusions}

The Landau-Lifshitz energy-momentum complex presents some singularities that
are determined by the metric structure. One of these singularities is $%
r\rightarrow 0$ which appears for any values of the phantom parameter $p$
and mass $M$ of the phantom black hole. The other singularities are the
roots of the second degree equation $6\arctan \bigl(\frac{r}{p}\bigr)%
p^{2}M+6\arctan \bigl(\frac{r}{p}\bigr)Mr^{2}+2p^{3}-3p^{2}M\pi +6pMr-3M\pi
r^{2}=0$.

As a conclusion, even it also yields positive values for the energy
distribution and gives physically acceptable results, the Landau-Lifshitz
energy-momentum complex is not the most suitable tool for the the
energy-momentum localization in the case of the phantom black hole. An
interesting future work lies in the calculation of the energy with the aid
of other energy-momentum complexes and the teleparallel equivalent to (GR).

\textbf{Acknowledgements}\ PKS acknowledges DST, New Delhi, India for
providing facilities through DST-FIST lab, Department of Mathematics, where
a part of this work was done. The authors also thank the referee for the
valuable suggestions, which improved the presentation of the obtained
results.

\end{document}